\documentclass[aps,prd,onecolumn,preprintnumbers,groupedaddress,showpacs,showkeys,
nofootinbib,amssymb]{revtex4}
\usepackage{graphicx}
\usepackage{epsfig}
\usepackage{bm,,color}
\usepackage{amssymb}
\usepackage{float}
\usepackage{amsmath}
\usepackage{cancel}

\allowdisplaybreaks[4]
\newcommand{\e}{\mathrm{e}}
\newcommand{\nn}{\nonumber \\}

\begin{document}

\title{Thermal effects and scalar modes in the cosmological propagation of gravitational waves}

\author{S.~Capozziello$^{1,2,3}$, S.~Nojiri$^{4,5}$, S.~D.~Odintsov$^{6,7,8}$}

\affiliation{\it $^1$ Dipartimento
di Fisica``E. Pancini'', Universit\`{a} di Napoli {}``Federico II''\\
$^2$INFN Sez. di Napoli, Compl. Univ. di Monte S. Angelo, Edificio G, 
Via Cinthia, I-80126, Napoli, Italy,\\
$^3$Scuola Superiore Meridionale, Largo S. Marcellino 10, I-80136, 
Napoli, Italy,\\
$^4$ Department of Physics,
Nagoya University, Nagoya 464-8602, Japan,\\
$^5$ Kobayashi-Maskawa Institute for the Origin of Particles and the universe, 
Nagoya University, Nagoya 464-8602, Japan,\\	
$^6$ICREA, Passeig Luis Companys,
23, 08010 Barcelona, Spain, \\
$^{7}$ Institute of Space Sciences (IEEC-CSIC)
C. Can Magrans s/n, 08193 Barcelona, Spain.
$^8$ Laboratory for Theoretical Cosmology,
Tomsk State University of Control Systems and Radioelectronics (TUSUR), 
634050 Tomsk, Russia.}

\date{\today}
\begin{abstract}
 We consider thermal effects in the  propagation of gravitational waves on a cosmological background.  In particular, we consider  scalar field cosmologies and study  gravitational  modes  near cosmological  singularities.
We point out 
that the contribution of thermal radiation can heavily affect the dynamics of gravitational 
waves giving enhancement or dissipation effects both at quantum and classical level.
These effects are considered both in General Relativity and in modified theories like $F(R)$ gravity which can be easily reduced to scalar-tensor cosmology. The possible detection and  disentanglement of standard and scalar  gravitational modes  on the stochastic background are also discussed.
\end{abstract}


\maketitle

\section{Introduction \label{SecI}}

The recent observations of gravitational waves and supermassive black holes can be 
considered as the main probes of General Relativity (GR) in its fundamental aspects which 
are: 1) the propagation of space-time perturbations, 2) the existence of singularities. Despite 
of these undeniable successes, several shortcomings affect GR because the whole 
phenomenology cannot be addressed in the framework of the Einstein picture. The theory is 
missing at ultraviolet scales because of the lack of a self-consistent theory of Quantum 
Gravity, and at infrared scales because it is not capable of encompassing clustering 
phenomena related to large-scale structure and the observed accelerated expansion of the 
cosmic fluid. These are generically dubbed as {\it dark matter} and {\it dark energy} but, up to now, no 
particle counterpart has been discovered to address them at fundamental level. 

In this perspective, extensions and modifications of GR are considered as a reliable way out 
of the above problems assuming that gravitational field has not been completely explored. 

These extensions come from effective theories on curved spacetimes 
\cite{Capozziello:2011et,Nojiri:2017ncd} or as alternative formulations like 
teleparallel gravity and its related models \cite{Cai:2015emx}.

A main role to test theories is played by cosmology because phenomena connected to the so 
called {\it dark side} can substantially affect structure formation and cosmic dynamics. Their 
equivalent geometric explanations could be a major step towards a comprehensive theory of 
gravity at all scales.

Specifically, the expansion of the universe is generated by cosmic fluids or, equivalently, by 
modified/extended gravity. 
Because the energy-momentum tensor of any fluid depends on the metric, 
the propagation of gravitational waves depends on what kind of model generates the 
expansion of the universe. Starting from this statement, gravitational waves can be a 
formidable tool to test cosmological models. A detailed discussion on this point is reported, 
for example, in \cite{Nojiri:2017hai,Bamba:2018cup}. 

In general, dynamical characteristics of gravitational waves can be the features probing a 
given theory of gravity \cite{Lombriser:2015sxa,Nakamura:2019yhf,Katsuragawa:2019uto,Lambiase,Bernal}. 
Specifically, speed, damping, 
dispersion, and oscillations of gravitational waves could be used to fix and reconstruct 
interactions into gravitational Lagrangian and then be a sort of roadmap inside the wide forest 
of competing theories of gravity. For example, further gravitational polarization modes, 
besides the two standard ones of GR, emerge when further degrees of freedom are 
considered into the theory 
\cite{Bogdanos:2009tn, Capozziello:2020xem, Capozziello:2020vil, DeLaurentis:2016jfs}. 
In general, as soon as 
modifications or extensions of GR are taken into account, scalar modes are present into 
dynamics.

Motivated by these considerations, it is possible to investigate the propagation of 
gravitational waves in various gravitational models. For example, 
in $F(T)$ extended teleparallel gravity \cite{Bamba:2013ooa}, in domain wall models 
\cite{Higuchi:2014bya}, in scalar-tensor and $F(R)$ gravity theories 
\cite{Capozziello:2017vdi}, in Chern-Simons Axion Einstein gravity 
\cite{Nojiri:2019nar,Nojiri:2020pqr} and in several media as in 
strong magnetic fields \cite{Bamba:2018cup} or in viscous fluids \cite{Brevik:2019yma}. 

Furthermore, the behavior of gravitational waves can be used to test past and future 
singularities and then contributes in their classification. 

Some previous results should  be mentioned in this perspective. For example, in \cite{Francaviglia, Felix}, the role of cosmological background is considered in the propagation of gravitational wave. In particular, the gauge invariance  and the conformal structure  are taken into account in order to fix both cosmological models and interferometric response. In  \cite{Francaviglia}, it is pointed out how the amplitude of a propagating gravitational  perturbation strictly depends on the  cosmological background and it is measurable, in principle, by the Sachs-Wolfe effect of the cosmic microwave background. On the other hand, this feature could constitute a fundamental tool to test modified gravity. In \cite{Vasilis1},  dynamics of gravitational waves in both late and early-time is considered for $F(R)$ gravity while, 
 in \cite{FelixSergey}, the behavior of gravitons in accelerated cosmology is taken into account. The main result of these studies is the indication that gravitons could be both a formidable feature to trace cosmic history from quantum gravity epoch up to late times and an unbiased approach to select cosmological models without imposing any dark component. This research trend could result   feasible also at interferometric level. In  \cite{Ricciardone1}, a detailed study on how cross-correlating astrophysical and cosmological gravitational wave backgrounds with the cosmic microwave is reported. Specifically, in 
 \cite{Ricciardone2}, the sensitivity of third-generation interferometers to extra polarizations in the stochastic gravitational wave background is discussed. The feeling is that self-consistent results  could be achieved soon matching together the VIRGO-LIGO interferometers, and the forthcoming LISA \cite{lisa} and Einstein Telescope \cite{ET}.
 
Another important issue is connected to thermal effects emerging during the cosmic 
evolution. As we will discuss in the next section, they are a very general feature related to 
the radius of the apparent horizon of a given singularity. At cosmological scales, such effects 
emerge with respect to the Hubble radius and then they can strongly affect the cosmic 
evolution, in particular, at early epochs or nearby singularities. In other words, thermal effects 
can dynamically affect the cosmological background and then the evolution of phenomena on 
it. 

In this paper, we want to investigate how thermal effects on various cosmological 
backgrounds affect the propagation of gravitational waves. In particular, we want to take into 
account such effects in GR, in modified theories of gravity and in presence of future 
singularities.

From the above point of view, considering the propagation of gravitational waves can be an 
important tool to discriminate among theories of gravity. In fact, as we will discuss, 
gravitational waves are a direct footprint of gravitational degrees of freedom and, if further 
modes emerge with respect to the two standard of GR, these are an important signature to 
extend or modify GR.
Furthermore, enhancement or dissipation of gravitational waves with thermal effects can 
probe the global cosmic evolution. In this perspective, a fine analysis of the gravitational 
stochastic background can be a further testbed for any theory of gravity.

The outline of the paper is the following. In Sec.~\ref{SecII}, 
we report a general discussion on thermal 
effects in cosmology. In particular, we take into account their relevance in classifying 
cosmological singularities and in generalized cosmologies where a scalar field is present. This 
last feature can be considered in a wide sense as characterizing extensions/modifications of 
GR. 

Sec.~\ref{SecIII} is devoted to the propagation of gravitational waves in a dynamical 
cosmological background. Here, we report how thermal radiation can affect the various 
components of gravitational waves.

Thermal corrections in quantum matter are discussed in Sec.~\ref{SecIV}. 
We want to show that such corrections can act both at quantum and classical level, so they 
have a relevant role also in quantum fluctuations of primordial epochs.

The specific effect of thermal radiation in early universe and nearby singularities is 
considered in Sec.~\ref{SecV}. The most relevant feature of this analysis is that thermal 
corrections can enhance or dissipate gravitational waves according to the 
signature and strength of the parameter $\alpha$ characterizing the thermal radiation.

Scalar modes in presence of thermal effects are discussed in Sec.~\ref{SecVI}. 
In particular, we demonstrate that scalar waves or compressional waves of cosmic fluids 
have similar effects in the cosmological evolution. According to this consideration, 
determining if the physical frame of a gravity theory is the Einstein or the Jordan one is 
crucial.

Sec.~\ref{SecVII} is devoted to the discussion of results and to the possibility of 
discriminating among the various contributions of gravitational radiation by the stochastic 
background. Here we give some qualitative evaluation on the fact that improving the 
sensitivity of interferometers could be a relevant issue for forthcoming experiments and 
observations on gravitational waves.

\section{Thermal effects in cosmology \label{SecII}}

In order to develop our considerations, let us start from a spatially flat 
Friedmann--Lema\^itre--Robertson--Walker (FLRW) universe defined by the metric
\begin{equation}
\label{JGRG14}
ds^2 = - dt^2 + a(t)^2 \sum_{i=1,2,3} \left(dx^i\right)^2\, .
\end{equation}
Here $a(t)$ is the cosmological scale factor.
Let us remind that the FLRW equations in General Relativity (GR), 
minimally coupled with a generic perfect fluid
with  pressure $p$ and  energy-density $\rho$ can be written as
\begin{equation}
\label{JGRG11}
\frac{3}{\kappa^2} H^2 = \rho \, ,\quad\quad
 - \frac{1}{\kappa^2}\left(3H^2 + 2\dot H\right)=p\, .
\end{equation}
Here $H\equiv \dot a/a$ is the Hubble parameter and $\kappa^2$ is the gravitational 
coupling.
When $H$ is large, the temperature of the universe becomes large and
we may expect the generation of thermal radiation as in the case 
of the Hawking radiation.
The Hawking temperature $T$ is proportional to the inverse of the radius
$r_\mathrm{H}$ of the apparent horizon and the radius $r_\mathrm{H}$ 
is proportional to the inverse of the Hubble rate $H$ \cite{Gibbons:1977mu}.
Therefore,  the temperature $T$ is proportional to the Hubble rate $H$.
As it is well known in statistical physics, the energy-density 
$\rho_\mathrm{t\_rad}$ of thermal radiation is 
proportional to the fourth power of temperature.
Then, when $H$ is large enough, we may assume that the energy-density of 
thermal radiation is given by
\begin{equation}
\label{BRHR1}
\rho_\mathrm{t\_ rad} = \alpha H^4 \, ,
\end{equation}
where $\alpha$ is a positive constant.
In this situation, the first of the FLRW Eqs.~(\ref{JGRG11}) can be modified 
taking into account the thermal radiation~\cite{Nojiri:2020sti,Ruggiero:2020piq}, that is, 
\begin{equation}
\label{BRHR2}
\frac{3}{\kappa^2} H^2 = \tilde\rho + \alpha H^4 \, .
\end{equation}
Here $\tilde\rho$ is the energy density of the cosmic fluid together with
the thermal radiation (\ref{BRHR1}). The fluid has to satisfy the conservation law, 
\begin{equation}
\label{consv}
\dot{\tilde\rho} + 3 H \left( \tilde\rho + \tilde p \right)=0\, ,
\end{equation}
coming from the contracted Bianchi identities. 
Here $\tilde p$ is the cosmic fluid pressure, also it considered together with thermal 
radiation. 

Combining the first FLRW Eq.~(\ref{BRHR2}) and 
the conservation law (\ref{consv}), we get the second FLRW equation, 
\begin{equation}
\label{TGW01}
 - \frac{1}{\kappa^2}\left(3H^2 + 2\dot H\right)
= \tilde p - \alpha \left( H^4 + \frac{4}{3} H^2 \dot H \right) \, , 
\end{equation}
where the effective pressure coming from thermal radiation is present.

When we can neglect the contribution of cosmic fluid with respect to 
the thermal radiation, that is $\tilde\rho = \tilde p = 0$, a non-trivial de Sitter solution is 
derived where $H$ is a constant, 
\begin{equation}
\label{BRHR4}
H^2 = H_\mathrm{crit}^2 \equiv \frac{3}{\kappa^2 \alpha} \, .
\end{equation}
If other matter contributions are present, the thermal radiation gives 
non-trivial effects. Specifically, it is worth stressing that thermal effects affect dynamics of 
future singularities that we are going to classify below.

\subsection{Thermal effects in future cosmological singularities}

It is well known that in the cosmic future, several kinds of space-time singularity can happen. 
Such singularities have been classified 
in Ref.~\cite{Nojiri:2005sx} (see also \cite{Odintsov:2018uaw} and \cite{Capozziello:2009hc}) 
as follows:
\begin{itemize}
\item Type I (``Big Rip'') Ref.~\cite{Caldwell:1999ew}:
This is a characteristic crushing type singularity, for which
as $t \to t_s$, the scale factor $a(t)$, the total effective pressure
$p_\mathrm{eff}$ and the total effective energy density $\rho_\mathrm{eff}$
diverge strongly, that is, $a \to \infty$, $\rho_\mathrm{eff} \to \infty$,
and $\left|p_\mathrm{eff}\right| \to \infty$.
This type of singularity is discussed in detail in 
Refs.~\cite{Capozziello:2009hc,Caldwell:2003vq,Nojiri:2003vn,Elizalde:2004mq,Faraoni:2001tq,
Singh:2003vx, Wu:2004ex,Sami:2003xv,Stefancic:2003rc,Chimento:2003qy,
Zhang:2005eg,Dabrowski:2006dd,Nojiri:2009pf,BeltranJimenez:2016dfc}.
\item Type II (``sudden''): This type of singularity is milder
than the Big Rip scenario, and it is also known as a pressure singularity,
firstly studied in Refs.~\cite{Barrow:2004xh,Nojiri:2004ip},
and later developed in \cite{Barrow:2004he,FernandezJambrina:2004yy,
BouhmadiLopez:2006fu,Barrow:2009df,BouhmadiLopez:2009jk,Barrow:2011ub,
Bouhmadi-Lopez:2013tua,Bouhmadi-Lopez:2013nma,
Chimento:2015gum,Cataldo:2017nck,Balcerzak:2012ae,Marosek:2018huv}.
Here, only the total effective pressure diverges as $t \to t_s$,
and the total effective energy density and the scale factor remain finite,
that is, $a \to a_s$, $\rho_\mathrm{eff} \to \rho_s$ and
$\left|p_\mathrm{eff}\right| \to \infty$.
\item Type III : In this type of singularity, both the total effective pressure
and the total effective energy density diverge as $t \to t_s$, but the scale
factor remains finite, that is, $a \to a_s$, $\rho_\mathrm{eff} \to \infty$ 
and $\left|p_\mathrm{eff}\right| \to \infty$.
Then Type III singularity is milder than Type I (Big Rip) but stronger than 
Type II (sudden).
\item Type IV : This type of singularity is the mildest from a phenomenological
point of view. 
It was discovered in Ref.~\cite{Nojiri:2004pf} and further investigated in
\cite{Nojiri:2005sx,Nojiri:2005sr,Barrow:2015ora,Nojiri:2015fra,
Nojiri:2015fia,Odintsov:2015zza,Oikonomou:2015qha,Kleidis:2017ftt}.
In this case, all the aforementioned physical quantities remain finite as
$t \to t_s$, that is, $a \to a_s$, $\rho_\mathrm{eff} \to 0$,
$\left|p_\mathrm{eff}\right| \to 0$,
but higher derivatives of the Hubble rate, $H^{(n)}$ $\left( n\geq 2 \right)$
diverge.
This singularity may be related with
the inflationary era, since the universe may smoothly pass through the
singularity without any catastrophic implications on the physical quantities.
As it was shown in \cite{Odintsov:2015gba}, the graceful exit from the
inflationary era may be triggered by this type of soft singularity.
\end{itemize}
Here, $\rho_\mathrm{eff}$ and $p_\mathrm{eff}$ are defined by
\begin{equation}
\label{IV}
\rho_\mathrm{eff} \equiv \frac{3}{\kappa^2} H^2 \, , \quad
p_\mathrm{eff} \equiv - \frac{1}{\kappa^2} \left( 2\dot H + 3 H^2
\right)\, ,
\end{equation}
and contain all pressures and densities that contribute as sources to the field equations. 
Then Eqs.~(\ref{IV}) show that for Type I and III singularities, $H$ diverges,
but for Type II and IV, $H$ is finite.
However, in Type II singularity, $\dot H$ diverges.

According to the considerations of Ref.~\cite{Nojiri:2020sti}, 
the thermal radiation usually makes the singularities 
{\it less singular}, that is, the Big Rip
(Type I) singularity or the Type III singularity transit to the Type II singularity. 

In the case of Big Rip singularity, which may be generated by ``phantom fields'', 
where the energy density $\rho$ behaves as 
\begin{equation}
\label{density}
\rho = \rho_0 a^{-3 \left( 1 + w \right)} \, ,
\end{equation}
with $w<-1$, the Hubble rate $H$ evolves as
\begin{equation}
\label{rip}
H \propto \frac{1}{t_\mathrm{Rip} - t} \, ,
\end{equation}
and $H$ diverges at $t=t_\mathrm{Rip}$, which is the Big Rip singularity. 
Near the Big Rip, the Hubble rate $H$ becomes large, and therefore 
the energy density (\ref{BRHR1}) of the thermal radiation becomes dominant. 
Then the non-trivial solution (\ref{BRHR4}) can be achieved. 
In such a solution, because $H$ goes to a constant, 
we might expect that the space-time becomes 
asymptotically de Sitter but it is not true.
Even in the de Sitter space-time, the scale factor $a(t)$ becomes larger and
larger as an exponential function of $t$, then the energy density (\ref{density}) 
of phantom fields should dominate finally.
The Hubble rate $H$ is, however, already larger than $H_\mathrm{crit}$, and then 
there is no solution in the FLRW equation, 
\begin{equation}
\label{BRHR2BB}
\frac{3}{\kappa^2} H^2 = \rho_0 a^{-3\left( 1 + w \right)} + \alpha H^4 \, .
\end{equation}
This situation tells us that the universe should end up at finite time with some kind of 
singularity.

By a more quantitative analysis, we find a maximum for the scale
factor $a$, that is
\begin{equation}
\label{BRHR6}
a \leq a_\mathrm{max} \equiv \left( \frac{9}{4 \kappa^4 \alpha \rho_0}
\right)^{- \frac{1}{3\left( 1 + w \right)}} \, , 
\end{equation}
and the Hubble rate $H$ behaves as 
\begin{align}
\label{BRHR11}
H \sim& \sqrt{ \frac{3}{2\alpha \kappa^2}}
\mp \frac{\sqrt{-3\left( 1 + w \right)}}{2} \left(
\sqrt{ \frac{3}{2\alpha \kappa^2}} \right)^\frac{3}{2}
\left( t_\mathrm{max} - t \right)^{\frac{1}{2}} \, , \nn
\dot H \sim& \mp \frac{\sqrt{-3\left( 1 + w \right)}}{4} \left(
\sqrt{ \frac{3}{2\alpha \kappa^2}} \right)^\frac{3}{2}
\left( t_\mathrm{max} - t \right)^{-\frac{1}{2}} \, .
\end{align}
Here we have assumed that $a=a_\mathrm{max}$ when $t=t_\mathrm{max}$.
Then, in the limit $t\to t_\mathrm{max}$, although $H$ is finite, 
$\dot H$ diverges.
Therefore the universe ends up with a Type II singularity at $t=t_\mathrm{max}$. 
Similarly, the Type III singularity also relaxes to become the 
Type II singularity. 

\subsection{Thermal effects in scalar-field cosmologies}

The above considerations can be extended to more general cosmologies. Let us take into 
account the contribution to $\tilde\rho$ and 
$\tilde p$ of some scalar field $\eta$ acting as a cosmic fluid. 
The Lagrangian density $\mathcal{L}_\eta$ of $\eta$ with a potential $V(\eta)$ 
has the following form:
\begin{equation}
\label{H7}
\mathcal{L}_\eta = - \frac{1}{2} \omega(\eta) \partial_\mu \eta
\partial^\mu \eta - V(\eta) \, .
\end{equation}
In the FLRW universe (\ref{JGRG14}), we can identify $\eta$ 
with the cosmological time $t$. 
Then the FLRW Eqs.~(\ref{BRHR2}) and (\ref{TGW01}) assume the forms:
\begin{align}
\label{TGW04}
\frac{3}{\kappa^2} H^2 =& \frac{1}{2} \omega(\eta) + V(\eta) + \alpha H^4 \, ,\nn
 - \frac{1}{\kappa^2}\left(3H^2 + 2\dot H\right)
=& \frac{1}{2} \omega(\eta) - V(\eta) - \alpha \left( H^4 + \frac{4}{3} H^2 \dot H \right) \, ,
\end{align}
which can be solved with respect to $\omega(\eta)$ and $V(\eta)$ as follows, 
\begin{equation}
\label{TGW05}
\omega(\eta) = - \frac{2}{\kappa^2} \dot H + \frac{4}{3}\alpha H^2 \dot H \, , \quad 
V(\eta)= \frac{1}{\kappa^2}\left(3H^2 + \dot H\right) 
 - \alpha \left( H^4 + \frac{2}{3} H^2 \dot H \right) \, .
\end{equation}
Therefore because $\eta=t$ and the r.h.s,'s of Eqs.~(\ref{TGW05}) are 
functions of $t$, if we choose $\omega(\eta)$ and $V(\eta)$ to satisfy 
Eq.~(\ref{TGW05}), any cosmology given by $H(t)$ can be realized. 
In the following, relations (\ref{TGW05}) will be considered for the 
propagation of cosmological gravitational waves. 

As an example, let us take into account a bouncing universe 
\begin{align}
\label{scale factor}
a(t) = a_0 \left(t^2 + t_0 \right)^n\, ,
\end{align}
where $a_0$, $t_0$, and $n$ are the model free parameters. 
When $t<0$, the universe is contracting, and, at $t=0$, 
the universe has a minimal size.
Then the universe starts to expand again for $t>0$. 
Eq.~(\ref{scale factor}) leads to the following Hubble rate
and its first derivative
\begin{align}
\label{bounce}
H(t) = \frac{2nt}{t^2 + t_0^2 }\, , \quad
\dot{H}(t) = -2n\frac{t^2 - t_0^2}{\left(t^2 + t_0^2 \right)^2}\, .
\end{align}
Therefore the Ricci scalar is found to be,
\begin{align}
R(t)=12H^2 + 6\dot{H}
=12n \left[\frac{(4n-1)t^2 + t_0^2}{\left(t^2 + t_0^2 \right)^2}\right] \, .
\label{ricci scalar}
\end{align}
Then Eqs (\ref{TGW05}) become 
\begin{align}
\label{TGW06}
\omega(\eta) =& -2n \left( - \frac{2}{\kappa^2} 
+ \frac{16 \alpha n^2t^2}{3 \left( t^2 + t_0^2 \right)^2}
\right) \frac{\left(t^2 - t_0^2\right)}{\left(t^2 + t_0^2 \right)^2} \, , \nn
V(\eta) =& \frac{\left( 12n^ - 2n \right) t^2 + 2n t_0^2}{\kappa^2\left(t^2 + t_0^2\right)^2}
 - \alpha \frac{4 n^2t^2 \left(
\left(12n^2 -4n \right) t^2 + 4n t_0^2\right)}{3\left(t^2 + t_0^2\right)^2} \, .
\end{align}
Note that $\frac{16 \alpha n^2t^2}{3 \left( t^2 + t_0^2 \right)^2}$ vanishes 
at $t=0$ and $t\to \pm \infty$ and the absolute value has the maximum 
value $\frac{4 \left| \alpha \right| n^2}{3 t_0^2}$ at $t=\pm t_0$
Then, if we choose $\frac{2}{\kappa^2}>\frac{4 \alpha n^2}{3 t_0^2}$, 
we find $- \frac{2}{\kappa^2} 
+ \frac{16 \alpha n^2t^2}{3 \left( t^2 + t_0^2 \right)^2} <0$. 
Therefore, when $t>t_0$ or $t<-t_0$, the scalar field $\eta$ 
is canonical but when $-t_0<t<t_0$, the scalar field has the wrong 
kinetic term: it is a phantom or a ghost field. In this example, it is clear that the thermal 
radiation assumes a key role in determining the evolution of the field and then of the 
universe.

\section{Gravitational Waves in a Dynamical Cosmological Background \label{SecIII}}

Let us consider now the propagation of gravitational waves in a dynamical cosmological 
background where thermal contributions are present. We want to show how these terms 
affect the evolution of gravitational waves.
First, we review the propagation of gravitational waves in a general medium.
Gravitational waves are derived as perturbations of Einstein field
equations.
In the Einstein equations, not only the curvature but also the energy-momentum tensor 
depends on the metric and therefore the variation of the energy-momentum tensor gives
a non-trivial contribution to the propagation of gravitational waves 
\cite{Nojiri:2017hai,Bamba:2018cup, Capozziello:2017xla}.

In general, the perturbed Einstein equations are given by
\begin{align}
\label{GRGW1}
0 =& \frac{1}{2\kappa^2}\left(- \frac{1}{2}\left(\nabla^{(0)}_\mu
\nabla^{(0)\, \rho} \delta g_{\nu\rho}
+ \nabla^{(0)}_\nu \nabla^{(0)\, \rho} \delta g_{\mu\rho} - \Box^{(0)} \delta g_{\mu\nu}
 - \nabla^{(0)}_\mu \nabla^{(0)}_\nu \left(g^{(0)\, \rho\lambda}\delta g_{\rho\lambda}\right)
\right. \right. \nn
& \left. - 2R^{(0)\, \lambda\ \rho}_{\ \ \ \ \ \nu\ \mu}\delta g_{\lambda\rho}
+ R^{(0)\, \rho}_{\ \ \ \ \ \mu}\delta g_{\rho\nu}
+ R^{(0)\, \rho}_{\ \ \ \ \ \nu}\delta g_{\rho\mu} \right) \nn
& \left. + \frac{1}{2} R^{(0)} \delta g_{\mu\nu}
+ \frac{1}{2}g^{(0)}_{\mu\nu} \left( -\delta g_{\rho\sigma} R^{(0)\, \rho\sigma}
+ \nabla^{(0)\, \rho} \nabla^{(0)\, \sigma} \delta g_{\rho\sigma}
 - \Box^{(0)} \left(g^{(0)\, \rho\sigma}\delta g_{\rho\sigma}\right) \right) \right)
+ \frac{1}{2} \delta T_{\mathrm{matter}\, \mu\nu} \, .
\end{align}
Here $\delta T_{\mathrm{matter}\, \mu\nu} \equiv 
\frac{\partial T_{\mathrm{matter}\, \mu\nu}}{\partial g_{\rho\sigma}}
\delta g_{\rho\sigma}$. 
For example, considering the scalar field $\eta$ in (\ref{H7}), we find
\begin{equation}
\label{H8}
T_{\mu\nu} = - \omega(\eta) \partial_\mu \eta \partial_\nu \eta
+ g_{\mu\nu} \mathcal{L}_\eta\, ,
\end{equation}
and therefore 
\begin{equation}
\label{Tphi2general}
\frac{\partial T_{\mu\nu}}{\partial g_{\rho\sigma}}
= \frac{1}{2} \left( \delta_\mu^{\ \rho} \delta_{\nu}^{\ \sigma}
+ \delta_\mu^{\ \sigma} \delta_{\nu}^{\ \rho} \right)
\left( - \frac{1}{2} g^{\eta\zeta} \omega(\eta) \partial_\eta \eta
\partial_\zeta \eta - V(\eta) \right)
+ \frac{1}{2} g_{\mu\nu} \omega(\eta) \partial^\rho \eta \partial^\sigma \eta
\, ,
\end{equation}
which we will use later. 

By multiplying the background metric $g^{(0)\, \mu\nu}$ with (\ref{GRGW1}), we obtain
\begin{equation}
\label{GRGW1B}
0 = \frac{1}{2\kappa^2}\left( \nabla^{(0)\, \sigma}\nabla^{(0)\, \rho} \delta g_{\sigma\rho}
 - \Box^{(0)} \left(g^{(0)\, \rho\lambda}\delta g_{\rho\lambda}\right)
+ \frac{1}{2} R^{(0)} \left(g^{(0)\, \rho\lambda}\delta g_{\rho\lambda}\right)
 - 2 \delta g_{\rho\sigma} R^{(0)\, \rho\sigma} \right)
+ \frac{1}{2} \delta T_\mathrm{matter} \, .
\end{equation}
We can choose the following gauge condition
\begin{equation}
\label{GRGW2}
0 = \nabla^{(0)\, \mu} \delta g_{\mu\nu} \, .
\end{equation}
Then Eq.~(\ref{GRGW1}) reduces to
\begin{align}
\label{GRGW3}
0 =& \frac{1}{2\kappa^2}\left(- \frac{1}{2}\left( - \Box^{(0)} \delta g_{\mu\nu}
 - \nabla^{(0)}_\mu \nabla^{(0)}_\nu \left(g^{(0)\, \rho\lambda}\delta g_{\rho\lambda}\right)
 - 2R^{(0)\, \lambda\ \rho}_{\ \ \ \ \ \nu\ \mu}\delta g_{\lambda\rho}
+ R^{(0)\, \rho}_{\ \ \ \ \ \mu}\delta g_{\rho\nu}
+ R^{(0)\, \rho}_{\ \ \ \ \ \nu}\delta g_{\rho\mu} \right) \right. \nn
& \left. + \frac{1}{2} R^{(0)} \delta g_{\mu\nu}
+ \frac{1}{2}g^{(0)}_{\mu\nu} \left( -\delta g_{\rho\sigma} R^{(0)\, \rho\sigma}
 - \Box^{(0)} \left(g^{(0)\, \rho\sigma}\delta g_{\rho\sigma}\right) \right) \right)
+ \frac{1}{2} \delta T_{\mathrm{matter}\, \mu\nu} \, ,
\end{align}
and Eq.~(\ref{GRGW1B}) to
\begin{equation}
\label{GRGW4}
0 = \frac{1}{2\kappa^2}\left(
 - \Box^{(0)} \left(g^{(0)\, \rho\lambda}\delta g_{\rho\lambda}\right)
+ \frac{1}{2} R^{(0)} \left(g^{(0)\, \rho\lambda}\delta g_{\rho\lambda}\right)
 - 2 \delta g_{\rho\sigma} R^{(0)\, \rho\sigma} \right)
+ \frac{1}{2} \delta T_\mathrm{matter} \, .
\end{equation}

By assuming the spatially flat FLRW space-time (\ref{JGRG14}),
we get
\begin{align}
\label{E2}
& \Gamma^t_{ij}=a^2 H \delta_{ij}\, ,\quad \Gamma^i_{jt}=\Gamma^i_{tj}=H\delta^i_{\ j}\, ,
\quad \Gamma^i_{jk}=\tilde \Gamma^i_{jk}\, ,\quad
R_{itjt}=-\left(\dot H + H^2\right)a^2\delta_{ij}\, ,\quad
R_{ijkl}= a^4 H^2 \left(\delta_{ik} \delta_{lj}
 - \delta_{il} \delta_{kj}\right)\, ,\nn
& R_{tt}=-3\left(\dot H + H^2\right)\, ,\quad
R_{ij}=a^2 \left(\dot H + 3H^2\right)\delta_{ij}\, ,\quad
R= 6\dot H + 12 H^2\, , \quad
\mbox{other components}=0\, .
\end{align}
Then $(t,t)$, $(i,j)$, $(t,i)$ components of (\ref{GRGW3}) assume the following forms:
\begin{align}
\label{GRGW5}
0 =& \frac{1}{2\kappa^2}\left( \frac{1}{2} \Box^{(0)} \delta g_{tt}
+ \frac{1}{2} \partial_t^2 \left(g^{(0)\, \rho\lambda}\delta g_{\rho\lambda}\right)
+ \frac{1}{2} \Box^{(0)} \left(g^{(0)\, \rho\sigma}\delta g_{\rho\sigma}\right) \right. \nn
& \left. - \frac{1}{2} \left(\dot H - H^2\right) \left(g^{(0)\, ij}\delta g_{ij}\right)
 - \frac{3}{2} \left(\dot H - H^2\right) \delta g_{tt} \right)
+ \frac{1}{2} \delta T_{\mathrm{matter}\, tt}
\, ,\\
\label{GRGW6}
0 =& \frac{1}{2\kappa^2}\left( \frac{1}{2} \Box^{(0)} \delta g_{ij}
+ \frac{1}{2} \left( \partial_i \partial_j - H \delta_{ij} \partial_t \right)
\left(g^{(0)\, \rho\lambda}\delta g_{\rho\lambda}\right)
 - \frac{1}{2} g^{(0)}_{ij} \Box^{(0)} \left(g^{(0)\, \rho\sigma}\delta g_{\rho\sigma}\right)
+ \frac{1}{2} \left( \dot H + H^2 \right) g^{(0)}_{ij} \delta g_{tt} \right. \nn
& \left. + 2 \left( \dot H + H^2 \right) \delta g_{ij}
 - \frac{1}{2} g^{(0)}_{ij} \left( \dot H + H^2 \right) \left(g^{(0)\, kl}\delta g_{kl}\right)
\right) + \frac{1}{2} \delta T_{\mathrm{matter}\, ij} \, ,\\
\label{GRGW7}
0 =& \frac{1}{2\kappa^2}\left( \frac{1}{2} \Box^{(0)} \delta g_{ti}
+ \frac{1}{2} \nabla^{(0)}_t \nabla^{(0)}_i \left(g^{(0)\, \rho\lambda}\delta g_{\rho\lambda}\right)
+ \left(2 \dot H + 4 H^2\right) \delta g_{ti} \right)
+ \frac{1}{2} \delta T_{\mathrm{matter}\, ti} \, .
\end{align}
Here $g^{(0)}_{ij}=a^2 \delta_{ij}$. 
For the scalar field $\eta$ in (\ref{H7}), by using (\ref{TGW05}) and (\ref{Tphi2general}), 
we find
\begin{align}
\label{TGW001}
\delta T^\eta_{tt} =& \left( - \frac{1}{\kappa^2}\left(3H^2 + \dot H\right)
+ \alpha \left( H^4 + \frac{2}{3} H^2 \dot H \right) \right) \delta g_{tt} \, , \nn
\delta T^\eta_{ij} =& \left( - \frac{1}{\kappa^2}\left(3H^2 + 2\dot H\right)
+ \alpha \left( H^4 + \frac{4}{3} H^2 \dot H \right) \right) \delta g_{ij}
+ \frac{1}{2} g^{(0)}_{ij} \left( - \frac{2}{\kappa^2} \dot H 
+ \frac{4}{3}\alpha H^2 \dot H \right) \delta g_{tt} \, , \nn
\delta T^\eta_{ti} =& \left( - \frac{1}{\kappa^2}\left(3H^2 + 2\dot H\right)
+ \alpha \left( H^4 + \frac{4}{3} H^2 \dot H \right) \right) \delta g_{ti} \, ,
\end{align}
where the thermal radiation contributions are clear and affect the evolution of gravitational 
waves.

\section{ Thermal Corrections in Quantum Matter  \label{SecIV}}

In order to find the explicit form of $\delta T_{\mathrm{matter}\, \mu\nu}$ 
in (\ref{GRGW1}) for the thermal radiation in (\ref{BRHR1}), 
we consider a real scalar field $\phi$ as the source of matter.
We deal with the scalar field as a quantum field at finite temperature.
In the case of high temperature or in the massless case, the scalar field
plays the role of radiation.
On the other hand, in the limit where the temperature is vanishing but the density
is finite, we obtain the dust, which can be considered as cold dark matter. 

In a curved space-time, the energy-momentum tensor of a real free scalar
field $\phi$ with mass $M$ is given by
\begin{equation}
\label{Tphi1}
T_{\mu\nu} = \partial_\mu \phi \partial_\nu \phi
+ g_{\mu\nu} \left( - \frac{1}{2} g^{\rho\sigma} \partial_\rho \phi
\partial_\sigma \phi - \frac{1}{2} M^2 \phi^2 \right) \, .
\end{equation}
In a flat background, we find
\begin{align}
\label{Tphi3b}
T_{tt} =& \rho = \frac{1}{2} \left( \pi^2 + \sum_{n=1,2,3}\left( \partial_n
\phi \right)^2
+ M^2 \phi^2 \right) \, , \nn
T_{ij} =& \partial_i \phi \partial_j \phi
+ \frac{1}{2} \delta_{ij} \left( \pi^2 - \sum_{n=1,2,3}
\left( \partial_n \phi \right)^2 - M^2 \phi^2 \right) \, .
\end{align}
Here $\pi=\dot\phi$ is the momentum conjugate to $\phi$.
We also obtain
\begin{equation}
\label{Tphi2}
\frac{\partial T_{\mu\nu}}{\partial g_{\rho\sigma}}
= \frac{1}{2} \left( \delta_\mu^{\ \rho} \delta_{\nu}^{\ \sigma}
+ \delta_\mu^{\ \sigma} \delta_{\nu}^{\ \rho} \right)
\left( - \frac{1}{2} g^{\eta\zeta} \partial_\eta \phi \partial_\zeta \phi
 - \frac{1}{2} M^2 \phi^2 \right)
+ \frac{1}{2} g_{\mu\nu} \partial^\rho \phi \partial^\sigma \phi \, ,
\end{equation}
which gets the following form in the flat background:
\begin{align}
\label{Tphi3}
\frac{\partial T_{tt}}{\partial g_{tt}} =& \frac{1}{4} 
\left( - \pi^2 - \sum_{n=1,2,3}\left( \partial_n \phi \right)^2 - M^2 \phi^2 \right) \, , \quad
\frac{\partial T_{tt}}{\partial g_{ij}} = - \frac{1}{2} \partial^i \phi \partial^j \phi \, , \quad 
\frac{\partial T_{ij}}{\partial g_{tt}} = - \frac{1}{2} \delta_{ij} \pi^2 \, , \nn
\frac{\partial T_{i0}}{\partial g_{j0}} =& 
\frac{\partial T_{i0}}{\partial g_{0j}} =\frac{\partial T_{0i}}{\partial g_{j0}} 
=\frac{\partial T_{0i}}{\partial g_{0j}} =
\frac{1}{4} \delta_i^j \left( \pi^2 - \sum_{n=1,2,3}\left( \partial_n \phi \right)^2 
 - M^2 \phi^2 \right) \, . \nn
\frac{\partial T_{ij}}{\partial g_{kl}} =& \frac{1}{4} \left( \delta_i^k
\delta_j^l + \delta_i^l \delta_j^k \right)
\left( \pi^2 - \sum_{n=1,2,3}\left( \partial_n \phi \right)^2 - M^2 \phi^2 \right)
+ \frac{1}{2} \delta_{ij} \partial^k \phi \partial^l \phi \, .
\end{align}
The thermal expectation values of other components vanish. 
Let us now evaluate the quantities in (\ref{Tphi3}) at
the finite temperature $T$.
In order to define the situation, we assume that the
three-dimensional space is the square box where the lengths of the edges are $L$ and we 
impose a periodic boundary condition on the scalar field $\phi$.
Then the momentum $\bm{k}$ is given by
\begin{equation}
\label{box1}
\bm{k} = \frac{2\pi}{L} \bm{n}\, , \quad \bm{n} =\left( n_x, n_y, n_z \right) \, .
\end{equation}
Here $n_x$, $n_y$, and $n_z$ are integers.
If we define,
\begin{equation}
\label{scalar8B}
\phi\left(\bm{x}\right)
\equiv \frac{1}{L^\frac{3}{2}} \sum_{\bm{n}}
\e^{i\frac{2\pi \bm{n} \cdot \bm{x}}{L}} \phi_{\bm{n}} \, , \quad
\pi\left(\bm{x}\right)
\equiv \frac{1}{L^\frac{3}{2}} \sum_{\bm{n}}
\e^{i\frac{2\pi \bm{n} \cdot \bm{x}}{L}} \pi_{\bm{n}} \, ,
\end{equation}
we find
\begin{equation}
\label{scalar9b}
\int d^3 x \phi\left(\bm{x}\right)^2
= \sum_{\bm{n}} \phi_{-\bm{n}} \phi_{\bm{n}}\, , \quad
\int d^3 x \pi\left(\bm{x}\right)^2
= \sum_{\bm{n}} \pi_{-\bm{n}} \pi_{\bm{n}}\, ,
\end{equation}
and the Hamiltonian is given by
\begin{equation}
\label{scalar1b}
H = \frac{1}{2} \sum_{\bm{n}} \left( \pi_{-\bm{n}} \pi_{\bm{n}}
+ E_{\bm{n}}^2\phi_{-\bm{n}} \phi_{\bm{n}}\right) \, , \quad 
E_{\bm{n}} \equiv
\sqrt{ \frac{\left(2\pi\right)^2 \bm{n}\cdot \bm{n}}{L^2} + M^2 } \, .
\end{equation}
Here $\pi_{\bm{k}}$ and $\phi_{\bm{l}}$ satisfy the following commutation
relation,
\begin{equation}
\label{scalar2b}
\left[ \pi_{\bm{n}}, \phi_{\bm{n'}} \right] = -i \delta_{\bm{n} + \bm{n'}, 0} \, .
\end{equation}
We now define the creation and annihilation operators $a^\pm_{\bm{n}}$ by
\begin{equation}
\label{scalar3b}
a^\pm_{\bm{n}}
= \frac{1}{\sqrt{2}} \left( \frac{\pi_{\bm{n}} }{\sqrt{E_{\bm{n}}}}
\pm i \sqrt{E_{\bm{n}}}\phi_{\bm{n}} \right) \, .
\end{equation}
We have to note that $\left( a^\pm_{\bm{n}} \right)^\dagger = a^\mp_{-\bm{n}}$ because
${\pi_{\bm{n}}}^\dagger = \pi_{-\bm{n}}$ and ${\phi_{\bm{n}}}^\dagger = \phi_{-\bm{n}}$.
The operators $a^\pm_{\bm{n}}$ satisfy the following commutation relations,
\begin{equation}
\label{scalar4b}
\left[ a^-_{\bm{n}}, a^+_{\bm{n'}} \right] = \delta_{\bm{n} + \bm{n'},0} \, ,
\quad \left[ a^\pm_{\bm{n}}, a^\pm_{\bm{n'}} \right] = 0 \, .
\end{equation}
Eqs.~(\ref{scalar3b}) can be solved with respect to $\pi_{\bm{n}}$
and $\phi_{\bm{n}}$ as follows,
\begin{equation}
\label{scalar5b}
\phi_{\bm{n}} = \frac{1}{i \sqrt{2E_{\bm{n}}}} \left( a^+_{\bm{n}} - a^-_{\bm{n}} \right) \, ,
\quad \pi_{\bm{n}} = \sqrt{\frac{E_{\bm{n}}}{2}}
\left( a^+_{\bm{n}} + a^-_{\bm{n}} \right) \, .
\end{equation}
Hamiltonian (\ref{scalar1b}) can be rewritten as
\begin{equation}
\label{scalar6b}
H = \sum_{\bm{n}} E_{\bm{n}}\left( a^+_{-\bm{n}} a^-_{\bm{n}} + \frac{1}{2} \right) \, .
\end{equation}
Let us now neglect the zero-point energy,
\begin{equation}
\label{scalar7b0}
H \to \tilde H = \sum_{\bm{n}} E_{\bm{n}} a^+_{-\bm{n}} a^-_{\bm{n}} \, .
\end{equation}
We define the number operator by
\begin{equation}
\label{number}
N \equiv \sum_{\bm{n}} a^+_{-\bm{n}} a^-_{\bm{n}} \, .
\end{equation}
Then we find the following expression of the partition function,
\begin{equation}
\label{scalar7b}
Z (\beta,\mu) = \mathrm{tr} \e^{ - \beta \tilde H - i \mu N }
= \e^{ - \sum_{\bm{n}}
\ln \left(1 - \e^{ - \beta E_{\bm{n}} - i\mu }\right)} \, .
\end{equation}
Here $\beta=\frac{1}{k_\mathrm{B}T}$ with the Boltzmann constant
$k_\mathrm{B}$ and $\mu$ is the chemical potential.
Then we find the thermal average of the operator
$a^+_{\bm{m}} a^-_{\bm{n}}$ is given as follows,
\begin{equation}
\left< a^+_{\bm{m}} a^-_{\bm{n}} \right>_{T,\mu}
= - \delta_{\bm{m} + \bm{n},0}
\frac{1}{\beta} \frac{\partial \ln Z (\beta,\mu) }{\partial E_{\bm{n}}}
= \delta_{\bm{m} + \bm{n},0}
\frac{\e^{ - \beta E_{\bm{n}} - i\mu }}{1 - \e^{ - \beta E_{\bm{n}} - i\mu }} \, .
\end{equation}
Then by a normal ordering, we have 
\begin{align}
\label{operators} 
: \pi^2 : =& \frac{1}{L^3} \sum_{\bm{m},\bm{n}}
\e^{i\frac{2\pi \left(\bm{m} + \bm{n}\right)\cdot \bm{x}}{L}} : \pi_{\bm{m}} \pi_{\bm{n}} 
= \frac{1}{L^3} \sum_{\bm{m},\bm{n}}
\e^{i\frac{2\pi \left(\bm{m} + \bm{n}\right)\cdot \bm{x}}{L}} \sqrt{ E_{\bm{m}} E_{\bm{n}}} 
a^+_{\bm{m}} a^-_{\bm{n}} \, , \nn
: \phi^2 : =& \frac{1}{L^3} \sum_{\bm{m},\bm{n}}
\e^{i\frac{2\pi \left(\bm{m} + \bm{n}\right)\cdot \bm{x}}{L}} : \phi_{\bm{m}} \phi_{\bm{n}} : 
= \frac{1}{L^3} \sum_{\bm{m},\bm{n}} \frac{
\e^{i\frac{2\pi \left(\bm{m} + \bm{n}\right)\cdot \bm{x}}{L}}}{\sqrt{ E_{\bm{m}} E_{\bm{n}}}} 
a^+_{\bm{m}} a^-_{\bm{n}} \, , \nn
: \partial^k \phi \partial^l \phi : =& \frac{1}{L^3} \sum_{\bm{m},\bm{n}}
\e^{i\frac{2\pi \left(\bm{m} + \bm{n}\right)\cdot \bm{x}}{L}} : \phi_{\bm{m}} \phi_{\bm{n}} : 
= \frac{1}{L^3} \sum_{\bm{m},\bm{n}} \frac{
\e^{i\frac{2\pi \left(\bm{m} + \bm{n}\right)\cdot \bm{x}}{L}}}{\sqrt{ E_{\bm{m}} E_{\bm{n}}}} 
m^k n^l a^+_{\bm{m}} a^-_{\bm{n}} \, ,
\end{align}
and we obtain 
\begin{align}
\label{operators2} 
\left< : \pi^2 : \right>_T 
=& \frac{1}{L^3} \sum_{\bm{n}} 
\frac{E_{\bm{n}} \e^{ - \beta E_{\bm{n}} - i\mu}}{1 - \e^{ - \beta E_{\bm{n}} - i\mu}} \, , 
\quad \left< : \phi^2 : \right>_T = \frac{1}{L^3} \sum_{\bm{n}} 
\frac{\e^{ - \beta E_{\bm{n}} - i\mu}}
{E_{\bm{n}} \left(1 - \e^{ - \beta E_{\bm{n}} - i\mu}\right)} \, , \nn
& \left< : \partial^k \phi \partial^l \phi : \right>_T = \frac{1}{3 L^3} \delta^{kl} 
\sum_{\bm{n}} \frac{\left(2\pi\right)^2 \bm{n}\cdot\bm{n} \e^{ - \beta E_{\bm{n}} - i\mu}}
{E_{\bm{n}} \left(1 - \e^{ - \beta E_{\bm{n}} - i\mu}\right)} \, .
\end{align}
In particular, we find 
\begin{equation}
\label{operators2BB} 
\left< : \left( \pi^2 - \sum_{n=1,2,3}\left( \partial_n \phi \right)^2 - M^2 \phi^2 \right)
: \right>_T=0 \, .
\end{equation}
Considering the operators in (\ref{Tphi3}), the thermal expectation values are given by 
\begin{align}
\label{scalar11D}
\left< : \frac{\partial T_{tt}}{\partial g_{tt}} : \right>_T
=& - \frac{1}{2L^3} \sum_{\bm{n}} 
\frac{E_{\bm{n}} \e^{ - \beta E_{\bm{n}} - i\mu}}{1 - \e^{ - \beta E_{\bm{n}} - i\mu}} \, , \nn
\left< : \frac{\partial T_{tt}}{\partial g_{ij}} : \right>_T
=& - \frac{\delta^{ij}}{6 L^3} 
\sum_{\bm{n}} \left\{ \frac{\left(2\pi\right)^2 \bm{n}\cdot\bm{n}}{L^2 E_{\bm{n}}} \right\}
\frac{\e^{ - \beta E_{\bm{n}} - i\mu}}
{\left(1 - \e^{ - \beta E_{\bm{n}} - i\mu}\right)} \, , \nn
\left< : \frac{\partial T_{ij}}{\partial g_{tt}} : \right>_T
=& - \frac{\delta_{ij}}{2L^3} \sum_{\bm{n}} 
\frac{E_{\bm{n}} \e^{ - \beta E_{\bm{n}} - i\mu}}{1 - \e^{ - \beta E_{\bm{n}} - i\mu}} \, , \nn
\left< : \frac{\partial T_{i0}}{\partial g_{j0}} : \right>_T
=& \left< : \frac{\partial T_{i0}}{\partial g_{0j}} : \right>_T
=\left< : \frac{\partial T_{0i}}{\partial g_{j0}} : \right>_T
=\left< : \frac{\partial T_{0i}}{\partial g_{0j}} : \right>_T = 0 \, . \nn
\left< : \frac{\partial T_{ij}}{\partial g_{kl}} : \right>_T
=& \frac{\delta_{ij} \delta^{kl}}{6L^3} \sum_{\bm{n}} \left\{
\frac{\left(2\pi\right)^2 \bm{n}\cdot\bm{n} }{L^2 E_{\bm{n}}} \right\}
\frac{\e^{ - \beta E_{\bm{n}} - i\mu }}{1 - \e^{ - \beta E_{\bm{n}} - i\mu }} \, .
\end{align}
In the limit of $L\to \infty$, we obtain
\begin{align}
\label{scalar11}
\left< : \frac{\partial T_{tt}}{\partial g_{tt}} : \right>_T
=& - \frac{1}{4\pi^2} \int_0^\infty dk \frac{\left( k^2 \sqrt{k^2 + M^2}\right) 
\e^{ - \beta \left( k^2 + M^2 \right)^\frac{1}{2} 
 -i\mu }}{1 - \e^{ - \beta \left( k^2 + M^2 \right)^\frac{1}{2} -i\mu }} \, . \nn
\left< : \frac{\partial T_{tt}}{\partial g_{ij}} : \right>_T
=& - \frac{\delta^{ij}}{12\pi^2} \int_0^\infty dk
\frac{k^4}{\sqrt{k^2 + M^2}}
\frac{\e^{ - \beta \left( k^2 + M^2 \right)^\frac{1}{2} -i\mu }}
{1 - \e^{ - \beta \left( k^2 + M^2 \right)^\frac{1}{2} -i\mu }} \, . \nn
\left< : \frac{\partial T_{ij}}{\partial g_{tt}} : \right>_T
=& - \frac{\delta_{ij}}{4\pi^2} \int_0^\infty dk 
\frac{\left( k^2 \sqrt{k^2 + M^2}\right) 
\e^{ - \beta \left( k^2 + M^2 \right)^\frac{1}{2} -i\mu }}
{1 - \e^{ - \beta \left( k^2 + M^2 \right)^\frac{1}{2} -i\mu }} \, . \nn
\left< : \frac{\partial T_{ij}}{\partial g_{kl}} : \right>_T
=& \frac{\delta_{ij} \delta^{kl}}{12\pi^2} \int_0^\infty dk
\frac{k^4}{\sqrt{k^2 + M^2}}
\frac{\e^{ - \beta \left( k^2 + M^2 \right)^\frac{1}{2} -i\mu }}
{1 - \e^{ - \beta \left( k^2 + M^2 \right)^\frac{1}{2} -i\mu }} \, .
\end{align}
In massless case $M=0$, by putting the chemical potential $\mu=0$, we find 
\begin{align}
\label{scalar12B}
& \left< : \frac{\partial T_{tt}}{\partial g_{tt}} : \right>_{T,\, M=\mu=0}
= - \frac{C}{4\pi^2 \beta^4} \, , \quad 
\left< : \frac{\partial T_{tt}}{\partial g_{ij}} : \right>_{T,\, M=\mu=0}
= - \frac{C \delta^{ij}}{12\pi^2 \beta^4} \, , \quad 
\left< : \frac{\partial T_{ij}}{\partial g_{tt}} : \right>_{T,\, M=\mu=0}
= - \frac{C \delta_{ij}}{4\pi^2 \beta^4} \, , \nn
& \left< : \frac{\partial T_{ij}}{\partial g_{kl}} : \right>_{T,\, M=\mu=0}
= \frac{C \delta_{ij} \delta^{kl}}{12\pi^2 \beta^4} \, , \quad 
C \equiv \int_0^\infty ds \frac{s^3 \e^{ - s }} {1 - \e^{ - s }} 
= \frac{\pi^4}{15} \, .
\end{align}
By using (\ref{Tphi3b}), we also find
\begin{align}
\label{scalar14}
\left< \rho \right>_T=& \frac{1}{2\pi^2} \int_0^\infty dk \frac{ k^2
\left( k^2 + M^2 \right)^\frac{1}{2}
\e^{ - \beta \left( k^2 + M^2 \right)^\frac{1}{2} -i\mu }}
{1 - \e^{ - \beta \left( k^2 + M^2 \right)^\frac{1}{2} -i\mu }} \, , \nn
\left< T_{ij} \right>_T = \delta_{ij} \left< p \right>_T =&
\frac{\delta_{ij}}{6 \pi^2 } \int_0^\infty d k \frac{ k^4
\e^{ - \beta \left( k^2 + M^2 \right)^\frac{1}{2} -i\mu }}
{\left( k^2 + M^2 \right)^\frac{1}{2}
\left( 1 - \e^{ - \beta \left( k^2 + M^2 \right)^\frac{1}{2} -i\mu }
\right)} \, .
\end{align}
In the massless limit $M\to 0$, by putting $\mu=0$, we get
\begin{equation}
\label{scalar15-00}
\left< \rho \right>_{T,\, M=\mu=0}
= 3 \left< p \right>_{T,\, M=\mu=0}
= \frac{C}{4\pi^2 \beta^4} \, ,
\end{equation}
which are very well-known results by the standard 
statistical physics. 
Comparing with (\ref{BRHR1}), that is, 
$\rho_\mathrm{t\_ rad}=\left< \rho \right>_{T,\, M=\mu=0}$, we find 
\begin{equation}
\label{TGW1}
C= 4\pi^2 \beta^4 \alpha H^4 \, .
\end{equation}
Then in the FLRW universe (\ref{JGRG14}), we have
\begin{align}
\label{TGW2}
\frac{\partial T^\mathrm{t\_ rad}_{tt}}{\partial g_{tt}}\equiv & 
\left< : \frac{\partial T_{tt}}{\partial g_{tt}} : \right>_{T,\, M=\mu=0}
= - \alpha H^4 \, , \quad 
\frac{\partial T^\mathrm{t\_ rad}_{tt}}{\partial g_{ij}} \equiv 
a^{-2} \left< : \frac{\partial T_{tt}}{\partial g_{ij}} : \right>_{T,\, M=\mu=0}
= - \frac{\alpha H^4}{3 a^2}\delta^{ij} \, , \nn
\frac{\partial T^\mathrm{t\_ rad}_{ij}}{\partial g_{tt}} \equiv& 
a^2 \left< : \frac{\partial T_{ij}}{\partial g_{tt}} : \right>_{T,\, M=\mu=0}
= - \alpha H^4 a^2 \delta_{ij} \, , \quad 
\frac{\partial T^\mathrm{t\_ rad}_{ij}}{\partial g_{kl}} \equiv 
\left< : \frac{\partial T_{ij}}{\partial g_{kl}} : \right>_{T,\, M=\mu=0}
= \frac{\alpha H^4}{3} \delta_{ij} \delta^{kl} \, , \nn
& \mbox{other components}=0 \, ,
\end{align}
and therefore
\begin{equation}
\label{TGW3}
\delta T^\mathrm{t\_ rad}_{tt} = - \alpha H^4 \left( \delta g_{tt} 
+ \frac{1}{3} g^{(0)\, ij} \delta g_{ij} \right) \, , \quad 
\delta T^\mathrm{t\_ rad}_{ij} = - \alpha H^4 g^{(0)}_{ij} \left( \delta g_{tt} 
 - \frac{1}{3} g^{(0)\, kl} \delta g_{kl} \right) \, .
\end{equation}
Because there can be many components contributing to the thermal radiation, 
we keep $\alpha$ as an unfixed parameter. This result clearly points out how, in principle, 
thermal radiation affect gravitational waves evolution in a cosmological background.

\section{Enhancement and Dissipation of Gravitational Waves with Thermal Effects 
\label{SecV}}

Let us now investigate the propagation of gravitational massless spin-two modes, where
\begin{equation}
\label{V9}
\delta g_{tt} = \delta g_{it}=\delta g_{ti}=0\, , \quad g^{(0)\, ij} \delta g_{ii}=0 \, , 
\end{equation}
Then Eqs.~(\ref{TGW3}) tell us that there is no direct 
affect in the propagation of gravitational waves from the thermal radiation. 
On the other hand, we see that Eqs.~(\ref{TGW001}) have the following form
\begin{equation}
\label{TGW4}
\delta T^\eta_{tt} = \delta T^\eta_{ti} = 0 \, , \quad 
\delta T^\eta_{ij} = \left( - \frac{1}{\kappa^2}\left(3H^2 + 2\dot H\right)
+ \alpha \left( H^4 + \frac{4}{3} H^2 \dot H \right) \right) \delta g_{ij} \, .
\end{equation}
The terms including $\alpha$ give the footstamps of the thermal radiation although 
Eqs.~(\ref{TGW3}) do not give any direct affect. 

Furthermore, combining (\ref{V9}) with the gauge condition (\ref{GRGW2}), we find that 
there is no longitudinal mode being
\begin{equation}
\label{TGW5}
0 = \partial^i \delta g_{ij} \, .
\end{equation}
Eqs.~(\ref{GRGW5}) and (\ref{GRGW7}) are trivially satisfied and Eq.~(\ref{GRGW6}) 
has the following forms, 
\begin{align}
\label{TGW6}
0 =& \frac{1}{2\kappa^2}\left( \frac{1}{2} \Box^{(0)} \delta g_{ij} 
+ 2 \left( \dot H + H^2 \right) \delta g_{ij} \right) 
+ \frac{1}{2} \delta T^\eta_{ij} \nn 
=& \frac{1}{4\kappa^2} \left( - \partial_t^2 \delta g_{ij} + 4 H \partial_t \delta g_{ij}
+ \left( 2 \dot H - 4 H^2 \right) \delta g_{ij} + a^{-2} \partial_k^2 \delta g_{ij} \right) 
+ \frac{\alpha}{2} \left( H^4 + \frac{4}{3} H^2 \dot H \right) \delta g_{ij} \, .
\end{align}
By writing $\delta g_{ij} = \e^{i\bm{k}\cdot\bm{x}} h_{ij}(t)$, 
Eq.~(\ref{TGW6}) can be written as 
\begin{equation}
\label{TGW7}
0 = \frac{1}{4\kappa^2} \left( - {\ddot h}_{ij} + 4 H {\dot h}_{ij}
+ \left( 2 \dot H - 4 H^2 \right) h_{ij} - a^{-2} k^2 h_{ij} \right) 
+ \frac{\alpha}{2} \left( H^4 + \frac{4}{3} H^2 \dot H \right) h_{ij} \, ,
\end{equation}
which is the evolution equation of the amplitude $h_{ij}$.
Here $k^2 \equiv \bm{k}\cdot\bm{k}$. The contribution of thermal term is clear and affects 
the evolution of the gravitational wave amplitude.

\subsection{The behavior of gravitational waves near the singularities}

Starting from the above results, we are able to study the behavior of  gravitational waves 
near the Type II singularity and the Big Rip (the Type I) singularity. 

Let us take into account the Type II singularity near $t=t_\mathrm{max}$ 
as in Eqs.~(\ref{BRHR11}), that is
\begin{equation}
\label{TGW8}
H = H_0 + H_1 \left( t_\mathrm{max} - t \right)^{\frac{1}{2}} + \cdots \, , 
\end{equation}
and therefore 
\begin{equation}
\label{TGW9}
\dot H = - \frac{H_1}{2} \left( t_\mathrm{max} - t \right)^{- \frac{1}{2}} + \cdots \, , 
\quad a=a_0 \e^{ - H_0 \left( t_\mathrm{max} - t \right) - \frac{2}{3} H_1 
\left( t_\mathrm{max} - t \right)^{\frac{3}{2}} 
+ \cdots} \, .
\end{equation}
We can also assume the behavior
\begin{equation}
\label{TGW10}
h_{ij} \propto \e^{f(t) + g(t) \left( t_\mathrm{max} - t \right)^{\frac{3}{2}}} \, .
\end{equation}
The functions $f(t)$ and $g(t)$ can be Taylor expanded 
 around $t=t_\mathrm{max}$, that is 
\begin{align}
\label{TGW11}
f(t) =& f_0 + f_1 \left( t_\mathrm{max} - t \right) 
+ \frac{1}{2} f_2 \left( t_\mathrm{max} - t \right)^2 + \cdots \, , \nn
g(t) =& g_0 + g_1 \left( t_\mathrm{max} - t \right) 
+ \frac{1}{2} g_2 \left( t_\mathrm{max} - t \right)^2 + \cdots \, . 
\end{align}
The parameter $f_0$ can be absorbed into the normalization of $h_{ij}$. 
Then by using Eq.~(\ref{TGW7}), we find, 
\begin{equation}
\label{TGW12}
0 = \frac{1}{4\kappa^2} \left( - \frac{3}{4} g_0 - H_1 \right) - \frac{1}{3}\alpha H_0^2 H_1 \, , 
\quad 0=\frac{1}{4\kappa^2} \left( - f_1^2 - f_2 - 4 H_0 f_1 - 4 H_0^2 - a_0^{-2} k^2 \right) 
+ \frac{\alpha}{2} H_0^4 \, ,
\end{equation}
which can be solved with respect to $g_0$ and $f_1$ as follows, 
\begin{equation}
\label{TGW13}
g_0 = - \frac{4}{3} H_1 \left( 1 + \frac{4\kappa \alpha}{3} H_0^2 \right) \, , \quad 
f_1 = - 2 H_0 \pm \sqrt{ - a_0^{-2} k^2 - \frac{f_2}{2} + 2 \kappa \alpha H_0^4} \, .
\end{equation}
In the case of the free scalar field, as shown in Eq.~(\ref{TGW1}), $\alpha$ is positive and 
therefore $g_0$ does not vanish and there appears non-analyticity in $h_{ij}$ at 
$t=t_\mathrm{max}$. 
If $\alpha$ is negative and satisfies the equation 
$1 + \frac{4\kappa \alpha}{3} H_0^2 =0$, $g_0$ vanishes and there is the possibility 
that the non-analyticity in $h_{ij}$ at $t=t_\mathrm{max}$ might not appear. 
When $- a_0^{-2} k^2 - \frac{f_2}{2} + 2 \kappa \alpha H_0^4<0$, $f_1$ becomes 
a complex number and therefore there occurs the oscillation as in the standard 
propagation of the gravitational wave. 
On the other hand, when $- a_0^{-2} k^2 - \frac{f_2}{2} + 2 \kappa \alpha H_0^4>0$, 
$f_1$ becomes a real number and therefore the oscillation does not occur. 
We have to note that $f_1$ is always negative. 
Therefore Eqs.~(\ref{TGW10}) and (\ref{TGW11}) tell us that the gravitational wave is 
always enhanced. 

Let us consider now the Big Rip (the Type I) singularity by putting $\alpha =0$ in 
(\ref{TGW7}), where
\begin{equation}
\label{BRGW1}
a\sim a_0 \left( t_s - t \right)^{-h} \, , \quad H \sim \frac{h}{t_s - t}\, .
\end{equation}
Here we assume $h$ is a positive constant. 
When we assume $h<2$, Eq.~(\ref{TGW7}) can be approximated as 
\begin{equation}
\label{BRGW2}
0 = - {\ddot h}_{ij} + \frac{4h}{t_s - t} {\dot h}_{ij}
+ \frac{2h - 4h^2 }{\left(t_s - t\right)} h_{ij} \, .
\end{equation}
By assuming $h_{ij} \propto \left(t_s - t\right)^n$, we find the following 
algebraic equation, 
\begin{equation}
\label{BRGW2BB}
0 = - n \left( n-1 \right) - 4h n + 2h - 4h^2 \, ,
\end{equation}
which can be solved with respect to $n$ as follows, 
\begin{equation}
\label{BRGW3}
n = -2 h \, , \ n= 1 - 2h \, .
\end{equation}
For the first mode, $h_{ij}$ is growing up but, if we consider $h_i^{\ j} = a^{-2} h_{ij}$, 
$h_i^{\ j}$ behaves as 
$h_i^{\ j} \to \mathrm{const.},$ and $\mathrm{const.}\times \left(t_s - t\right)$. 
Therefore there is no enhancement and there is also a dissipating mode. 
For the Type II case in (\ref{TGW9}), $a(t)$ goes to a finite constant when 
$t\to t_\mathrm{max}$ and therefore the qualitative behavior of $h_i^{\ j}$ does not change 
from $h_{ij}$. 
This tells us that the enhancement of the gravitational wave occurs near the Type II 
singularity but it could not occur near the Big Rip (Type I) singularity. 

\subsection{Gravitational waves in the early universe}

We may also consider the bouncing universe as in (\ref{bounce}). 
Because $H$ vanishes at the bouncing point $t=0$, we may approximate $H$ as 
$H=H_1 t$ $\left( H_1>0\right)$. 
Then Eq.~(\ref{TGW7}) can be approximated as 
\begin{equation}
\label{TGW14}
0 = - {\ddot h}_{ij} + 2 H_1 h_{ij} - a_0^{-2} k^2 h_{ij} \, ,
\end{equation}
whose solution is given by 
$h_{ij} \propto \e^{\pm i \sqrt{a_0^{-2} k^2 - 2 H_1}}$. 
Therefore in the high frequency modes, where $k^2 > 2 a_0^2 H_1$, the gravitational 
waves develop in a standard way with oscillations. 
On the other hand, for the low frequency modes, where $k^2 < 2 a_0^2 H_1$, the amplitude of 
the gravitational waves increases for $+$ signature and decreases for $-$ signature. 
Therefore the low frequency modes are enhanced or dissipated  depending on the initial 
conditions. 

During the inflationary epoch, where $H$ is almost constant, $H=H_0$ and 
$a=a_0 \e^{H_0 t}$, Eq.~(\ref{TGW7}) has the following form. 
\begin{equation}
\label{TGW15}
0 = \frac{1}{4\kappa^2} \left( - {\ddot h}_{ij} + 4 H_0 {\dot h}_{ij}
 - 4 H_0^2 h_{ij} - a_0^{-2} \e^{-2 H_0 t} k^2 h_{ij} \right) 
+ \frac{\alpha}{2} H_0^4 h_{ij} \, .
\end{equation}
Due to the expansion of the universe, the term including $\e^{-2 H_0 t}$ becomes very small 
and we can neglect it. 
Then, assuming $h_{ij} \propto \e^{\lambda t}$ with a constant $\lambda$, we find 
\begin{equation}
\label{TGW16}
0 = \frac{1}{4\kappa^2} \left( - \lambda^2 + 4 H_0 \lambda - 4 H_0^2 \right) 
+ \frac{\alpha}{2} H_0^4 \, ,
\end{equation}
which can be solved with respect to $\lambda$ as follows, 
\begin{equation}
\label{TGW17}
\lambda = 2 H_0 \pm H_0^2 \sqrt{\frac{\alpha}{2}} \, .
\end{equation}
Then we find $h_{ij} \propto \e^{\left( 2 H_0 \pm H_0^2 \sqrt{\frac{\alpha}{2}} \right)t}$ 
and $h_{ij} \propto \e^{\pm H_0^2 t \sqrt{\frac{\alpha}{2}} }$. 
If there is no thermal effect, that is $\alpha=0$, there is no enhancement 
or dissipation of the gravitational wave, but if we include the thermal effect, 
enhancement or dissipation occur.

\section{The propagation of Scalar Modes with Thermal Effects \label{SecVI}}

A similar discussion can be developed also in a generalized context where scalar modes are 
included.

Considering the model (\ref{H7}) describing the dynamics of the field $\eta$, we can take 
into account the propagation of scalar modes. 
Let us find  the expression of the effective mass of the scalar field. 
By redefining the scalar field $\eta$ as 
\begin{equation}
\label{SPr1}
\zeta = \int d\eta \sqrt{ \left| \omega(\eta) \right|} \, , 
\end{equation}
we rewrite the Lagrangian density $\mathcal{L}_\eta$ in (\ref{H7}), 
as follows \begin{equation}
\label{SPr2}
\mathcal{L}_\eta = - \frac{1}{2} \mathrm{sign}\left( \omega \left( \eta \right) \right) 
\partial_\mu \zeta \partial^\mu \zeta - V\left(\eta\left(\zeta\right)\right) \, .
\end{equation}
Here the $\mathrm{sign}$ function is defined as 
\begin{equation}
\label{SPr3}
\mathrm{sign} (x) \equiv \left\{ 
\begin{array}{cc}
1 & \mbox{if}\ x > 0 \\
 -1 & \mbox{if}\ x < 0 
\end{array} \right. \, .
\end{equation}
Then the square of mass $m$ is found to be 
\begin{align}
\label{SPr4B}
m^2 \equiv& \mathrm{sign}\left( \omega \left( \eta \right) \right)\frac{d^2 V}{d\zeta^2} \nn
=& \mathrm{sign}\left( \omega \left( \eta \right) \right)
\frac{1}{\sqrt{\left| \omega(\eta) \right|}} \frac{d}{d\eta} \left( 
\frac{1}{\sqrt{\left| \omega(\eta) \right|}} \frac{dV}{d\eta} \right) \nn
=& \frac{1}{\omega(\eta)} \frac{d^2V}{d\eta^2} 
 - \frac{\omega'(\eta)}{2\omega(\eta)^2} \frac{dV}{d\eta} \, .
\end{align}
Assuming $\eta=t$, by using Eqs.~(\ref{TGW05}), we find
\begin{align}
\label{SPr4}
m^2 =& \frac{1}{2}
\left( - \frac{2}{\kappa^2} \dot H + \frac{4}{3}\alpha H^2 \dot H \right)^{-2}
\left\{ \frac{2}{\kappa^4} \left( -12 {\dot H}^3 - 6 H \dot H \ddot H 
 - 2 \dot H\dddot H + {\ddot H}^2 \right) \right. \nn
& + \frac{2\alpha}{\kappa^2} \left( 24 H^2 {\dot H}^3 + 8 H^3 \dot H \ddot H 
+ \frac{8}{3} {\dot H}^4 + \frac{16}{3} H {\dot H}^2 \ddot H 
+ \frac{8}{3} H^2 \dot H \dddot H - \frac{4}{3} H^2 {\ddot H}^2 \right) \nn
& \left. + \alpha^2 \left( - \frac{64}{3} H^4 {\dot H}^3 - \frac{16}{3} H^5 \dot H \ddot H 
 - \frac{64}{9} H^3 {\dot H}^2 \ddot H + \frac{8}{9} H^4 {\ddot H}^2 \right) \right\} \, .
\end{align}
It is worth noticing that when 
$\omega(\eta)=- \frac{2}{\kappa^2} \dot H + \frac{4}{3}\alpha H^2 \dot H=0$, 
that is, when $\omega(\eta)$ changes its signature, $m^2$ diverges. 

In case of the Type II singularity (\ref{TGW8}), 
around the singularity $t=t_\mathrm{max}$, 
we find 
\begin{equation}
\label{SPr5}
m^2 \sim \frac{1}{16\left( t_\mathrm{max} - t \right)^2} \, ,
\end{equation}
which does not depend on $H_0$ nor $H_1$ and $m^2$ is positive 
and diverges at the singularity $t=t_\mathrm{max}$. 
Therefore just before the singularity, the scalar field oscillates very rapidly. 

The Big Rip (the Type I) singularity (\ref{BRGW1}) is recovered for 
$\alpha =0$. In this case, we have
\begin{equation}
\label{SPr6}
m^2 \sim - \frac{6h + 2}{\left( t_s - t \right)^2} \, .
\end{equation}
Because $m^2$ is negative and diverges at $t=t_s$, 
the amplitude of scalar field increases or decreases very rapidly. 

In the case of bouncing universe (\ref{scale factor}), 
as we find in (\ref{TGW06}), $\omega(\eta)$ vanishes at $t=\pm t_0$ 
and therefore $m^2$ diverges. 
If $m^2>$, which may depend on the parameters near $t= \pm t_0$, 
the scalar field oscillates very rapidly and if $m^2<0$, 
the amplitude of the scalar field increases or decreases very rapidly. 

In the inflationary era, where $H$ is almost constant $H\sim H_0$, 
$m^2$ in (\ref{SPr4}) has the following form 
\begin{align}
\label{SPr7}
m^2 =& \frac{1}{2}
\left( - \frac{2}{\kappa^2} + \frac{4}{3}\alpha H_0^2 \right)^{-2}
\left\{ \frac{2H_0^6}{\kappa^4} \left( -12 \epsilon_H^3 - 6 \epsilon_H^2 \eta_H 
 - 2 \epsilon_H \xi_H + \epsilon_H^2 \eta_H^2 \right) \right. \nn
& + \frac{2\alpha H_0^8}{\kappa^2} \left( 24 \epsilon_H^3 + 8 \epsilon_H^2 \eta_H
+ \frac{8}{3} \epsilon_H^4 + \frac{16}{3} \epsilon_H^3 \eta_H 
+ \frac{8}{3} \epsilon_H \xi_H - \frac{4}{3} \epsilon_H^2 \eta_H^2 \right) \nn
& \left. + \alpha^2 H_0^{10} \left( - \frac{64}{3} \epsilon_H^3 
 - \frac{16}{3} \epsilon_H^2 \eta_H - \frac{64}{9} \epsilon_H^3 \eta_H 
+ \frac{8}{9} \epsilon_H^2 \eta_H^2 \right) \right\} \, .
\end{align}
Here the slow roll parameters $\epsilon_H$, $\eta_H$, and $\xi_H$ are defined by
\begin{equation}
\label{SPr8}
\epsilon_H \equiv \frac{\dot H}{H^2}\, , \quad 
\eta_H \equiv \frac{\ddot H}{H\dot H}\, , \quad 
\xi_H \equiv \frac{\dddot H}{H^4} \, .
\end{equation}
Therefore the mass $m$ could be finite but the signature of $m^2$ 
depends on the details of inflation. 

The expanding universe can be realized by the perfect fluid. 
The perfect fluid also generates scalar waves, whose propagating velocity 
is known as the sound speed $c_s$, which is given by 
\begin{equation}
\label{SPr9}
c_s^2 = \frac{dp}{d\rho}\, .
\end{equation}
Therefore if we find the equation of state (EoS), we can find the speed. 
The energy density $\rho$ and the pressure $p$ are given in the FLRW 
Eqs.~(\ref{JGRG11}). 
The left hand sides of the FLRW equations are given by functions of 
the cosmological time $t$. 
Therefore if time is intrinsic in the expressions of 1st and 2nd FLRW equations, we can 
find the EoS as the ratio of pressure and energy density.
The FLRW equations also tell the $t$ dependence of the sound speed, as follows, 
\begin {equation}
\label{SPr10}
c_s^2 = \frac{\frac{dp}{dt}}{\frac{d\rho}{dt}}
= - 1 - \frac{\ddot H}{3 H \dot H} \, .
\end{equation}
Then we find the sound speed from the time dependence of the Hubble rate $H$. 

In case of the Type II singularity (\ref{TGW8}), we find
\begin{equation}
\label{SPr11}
c_s^2 \sim - \frac{1}
{6 H_0 \left( t_\mathrm{max} - t \right) }\,.
\end{equation}
Therefore $c_s^2$ diverges to negative infinity. 
Because $c_s^2$ is negative, the amplitude of the perfect fluid wave 
rapidly decreases or increases without oscillation. 
The behavior is much different from that in the scalar field in (\ref{SPr5}), 
where the scalar field oscillates very rapidly. 

In case of the Big Rip (the Type I) singularity (\ref{BRGW1}), we find 
\begin{equation}
\label{SPr12}
c_s^2 \sim - 1 - \frac{2}{3h} \, , 
\end{equation}
which is finite but because $c_s^2$ is negaitve, 
the amplitude of the perfect fluid wave 
decreases or increases exponentially without oscillation. 
Even in case of the scalar field in (\ref{SPr6}), 
the amplitude of the scalar field increases or decreases very rapidly 
but $m^2$ diverges in the case of scalar field, the increase or decrease 
is much more rapid. 

In case of the bouncing universe (\ref{scale factor}), the 
sound speed is given by 
\begin {equation}
\label{SPr13}
c_s^2 = - 1 - \frac{ -2n \frac{2t \left(t^2 + t_0^2 \right) 
 - 4t \left( t^2 - t_0^2 \right)}{\left(t^2 + t_0^2 \right)^3}}
{\frac{6nt}{t^2 + t_0^2 }
\left( -2n\frac{t^2 - t_0^2}{\left(t^2 + t_0^2 \right)^2} \right)}
= - 1 - \frac{- t^2 + 3 t_0^2}
{3n\left(t^2 - t_0^2\right)} \, ,
\end{equation}
which diverges at $t=t_0$ as  the mass $m^2$ of the scalar field. 
Then the propagation of the perfect fluid wave might be similar to 
that in the scalar field. 

In case of the inflation, $H\sim H_0$, we find
\begin {equation}
\label{SPr14}
c_s^2 = - 1 - \frac{\eta_H}{3}\, .
\end{equation}
which is finite but could be negative as long as $\eta_H$ is small enough. 
Therefore the perfect scalar wave does not propagate although the scalar wave 
can propagate. 

The above results tell us that the propagation of scalar modes depends on 
the mechanism which generates the expansion of the universe and, since thermal effects 
affect the effective mass \eqref{SPr4}, they have to be considered in the evolution.

\subsection{Scalar waves in modified gravity versus compressional waves of cosmic fluid}

Scalar modes can be achieved also taking into account modified theories of gravity, in 
particular higher-order theories in the curvature invariants. As a specific case, let us 
consider $F(R)$ gravity \cite{Capozziello:2002rd, Capozziello:2011et, Nojiri:2010wj, 
Maeda:1988ab, Nojiri:2003ft}. In this case, it is possible to compare scalar gravitational waves 
with the compressional waves generated by a cosmic fluid. 

The action of $F(R)$ gravity is given by 
\begin{equation}
\label{JGRG7}
S_{F(R)}= \int d^4 x \sqrt{-g} \left( \frac{F(R)}{2\kappa^2} 
+ \mathcal{L}_\mathrm{matter} \right)\, .
\end{equation}
Here $F(R)$ is a generic function of the Ricci scalar $R$ and $\mathcal{L}_\mathrm{matter}$ 
is the Lagrangian density of standard matter. 
It is well known that $F(R)$ gravity can be rewritten 
in a scalar-tensor form \cite{Capozziello:2011et, Nojiri:2010wj, Maeda:1988ab,Nojiri:2003ft}.
By introducing an auxiliary field $A$, the action (\ref{JGRG7}) of 
$F(R)$ gravity assumes the following form:
\begin{equation}
\label{JGRG21}
S=\frac{1}{2\kappa^2}\int d^4 x \sqrt{-g} \left\{F'(A)\left(R-A\right) 
+ F(A)\right\}\, .
\end{equation}
By the variation of $A$, one obtains $A=R$. Substituting $A=R$ into
the action (\ref{JGRG21}), one reproduces immediately the action in (\ref{JGRG7}). 
Furthermore, we can conformally rescale the metric in the following way, 
\begin{equation}
\label{JGRG22}
g_{\mu\nu}\to \e^\sigma g_{\mu\nu}\, ,\quad \sigma = -\ln F'(A)\, ,
\end{equation}
obtaining the action in the Einstein frame, that is 
\begin{align}
\label{JGRG23}
S_E =& \frac{1}{2\kappa^2}\int d^4 x \sqrt{-g} \left( R 
 - \frac{3}{2}g^{\rho\sigma}
\partial_\rho \sigma \partial_\sigma \sigma - V(\sigma)\right)\,,
\end{align}
with the effective potential 
\begin{align}
V(\sigma) =& \e^\sigma g\left(\e^{-\sigma}\right)
 - \e^{2\sigma} f\left(g\left(\e^{-\sigma}\right)\right) 
= \frac{A}{F'(A)} - \frac{F(A)}{F'(A)^2}\, .
\end{align}
Here $g\left(\e^{-\sigma}\right)$ is given by solving the equation
$\sigma = -\ln\left( 1 + f'(A)\right)=- \ln F'(A)$ as
$A=g\left(\e^{-\sigma}\right)$.
Due to the scale transformation (\ref{JGRG22}), a coupling
of the scalar field $\sigma$ with standard matter arises. 
The mass of $\sigma$ is given by
\begin{equation}
\label{JGRG24}
m_\sigma^2 \equiv \frac{3}{2}\frac{d^2 V(\sigma)}{d\sigma^2}
=\frac{3}{2}\left\{\frac{A}{F'(A)} - \frac{4F(A)}{\left(F'(A)\right)^2} 
+ \frac{1}{F''(A)}\right\}\, .
\end{equation}
Then in the framework of $F(R)$ gravity, a propagating scalar mode appears and the 
considerations in Sec.~\ref{SecIV} can be applied.

As an example, let us take into account the case of power-law $F(R) = f_0 R^m$. 
It is straightforward obtaining the exact cosmological solution, where the Hubble rate $H$ 
is given by 
\begin{equation}
\label{JGRG17}
H = \frac{-\frac{(m-1)(2m-1)}{m-2}}{t}\, .
\end{equation}
and the curvature scalar is 
\begin{equation}
\label{RA}
R=A= 12 H^2 + 6 \dot H 
= \frac{6m(4m - 5)(m-1)(2m-1)}{(m-2)^2t^2} \, .
\end{equation}
The same solution can be also obtained in the Einstein gravity coupled with 
the cosmic perfect fluid with a constant EoS parameter. 
In fact, modeling out the further degrees of freedom of $F(R)$ gravity as a perfect fluid, 
it is easy to obtain \cite{Capozziello:2019qlt}
\begin{equation}
\label{JGRG18}
w \equiv \frac{p}{\rho} =-\frac{6m^2 - 7m - 1}{3(m-1)(2m -1)}\, .
\end{equation}
Clearly, in the Einstein gravity, the propagating scalar mode does not appear because
the situation is different from that in $F(R)$ gravity. 
Instead of the massive scalar mode, 
a compressional wave, generated by the cosmic fluid, appears. In this case, 
the sound speed is constant as given from (\ref{SPr9}), 
\begin {equation}
\label{SPr15}
c_s^2 = \frac{dp}{d\rho} = w \, .
\end{equation}
When $w$ is positive, the compressional wave propagates but 
when $w$ is negative, the amplitude of the wave increases or decreases 
exponentially and therefore the wave does not propagate. 

Because we are considering the case $F(R) = f_0 R^m$, 
the mass $m_\sigma^2$ has the following form 
\begin{equation}
\label{JGRG24B}
m_\sigma^2 \equiv \frac{3}{2f_0 A^{m-2}}
\left( \frac{1}{m} - \frac{4}{m^2} + \frac{1}{m(m-1)}\right)
= \frac{3 \left( m - 2 \right)^2}{2m^2 (m-1) f_0 A^{m-2}}\, .
\end{equation}
As long as $F'(R)=mf_0 R^{m-1} = mf_0 A^{m-1} >0$ and $m_\sigma^2>0$, 
a propagating scalar mode appears.

As an example, let us consider the case $m\gg 1$. 
Here we find 
\begin {equation}
\label{SPr16}
w\sim -1 <0\, , \quad R=A \sim \frac{48m^2}{t^2}>0\, , \quad 
m_\sigma^2 \sim \frac{3t^{2\left( m - 2\right)}}{f_0 m \left( 48m^2 \right)^{m-2}} \, .
\end{equation}
If $f_0>0$, we find $F'(R)>0$ and $m_\sigma^2>0$, and therefore 
the massive scalar mode in the $F(R)$ propagates but because $c_s^2=w \sim -1 <0$, 
the compressional wave of the cosmic fluid does not propagate. 
Hence the expansion of the universe is identical between the $F(R)$ gravity and the 
Einstein gravity with the cosmic perfect fluid but, only in the $F(R)$ gravity, 
a propagating scalar mode appears. From an observational point of view, 
these feature can be extremely relevant because it is fixing, if detected, if the 
physical frame is the Jordan one (with $F(R)$ gravity) or the Einstein one (with perfect fluid). 
See also \cite{Capozziello:2006dj} for a discussion.

It is worth saying that, in various modified and extended theories of gravity, like $F(R)$, 
Brans-Dicke, scalar-tensor, or Gauss-Bonnet gravity, scalar modes appear and they  are often 
massive. 
The expansion of the background universe, generated by these models can be 
also generated by the Einstein gravity sourced by perfect fluids. 
The perfect fluid is characterized by the energy density $\rho$, the pressure $p$, and the 
EoS $p=f\left(\rho, \cdots\right)$. 
The effective energy density $\rho_\mathrm{eff}$, the effective pressure $p_\mathrm{eff}$ 
in the modified gravities satisfy an identical EoS 
$p_\mathrm{eff}=f\left(\rho_\mathrm{eff}, \cdots\right)$ in the 
homogeneous background. In other words, the effect of further gravitational degrees of 
freedom, related to modified gravity theories, can be  represented as perfect fluids if certain conditions are satisfied. See 
 \cite{Capozziello:2019qlt, Capozziello:2019wfi, Capozziello:2018ddp}. 
However, the point is that the propagation of  scalar modes or the fluid dynamics fix 
the frame and can be the physical signature discriminating between models.
 
If we consider inhomogeneous space-times, the behavior of modified gravity 
is different from that of Einstein gravity with the perfect fluid. 
For example, when we consider spherically symmetric background, the effective pressure 
in the radial direction is generally different from the effective pressure in the angular 
direction although the pressure in the perfect fluid should not depend on the direction. 
This tells us that, if we consider a perturbation on the homogeneous background like a wave, 
the behavior in modified gravity can be different from that in the Einstein gravity 
with the perfect fluid as discussed in this section. 

Then how we can distinguish scalar modes with respect to the compressional waves of a 
perfect fluid? 
As we have seen, a perfect fluid, where the EoS parameter is $w<0$, does not generate a 
compressional wave. Therefore if we find massive scalar waves, it could be an evidence for 
modified gravity. However, even in modified gravity, there are various models. 
In the case of $F(R)$ gravity, however, the coupling of massive scalar with matter is 
universal, that is, it does not depend on the kind of matter because the coupling appears by 
the rescaling of  metric (\ref{JGRG22}). This structure is rather characteristic of $F(R)$ 
gravity and it may give some clue to observationally discriminate $F(R)$ gravity with respect 
to other modified gravity models. 

\section{Discussion and conclusions \label{SecVII}}

 In this paper, we discussed thermal radiation effects affecting the propagation of 
cosmological gravitational waves both in the context of GR and in its  
modifications when scalar gravitational modes emerge. 
In particular, thermal effects have a main role in determining both the cosmological parameters and the evolution towards early and future singularities. The key  feature is the dependence of thermal radiation from 
the Hubble parameter   so that,  for large $H$, the temperature of the universe becomes large and
we may expect effects  as in the case 
of the Hawking radiation. As discussed, 
the Hawking temperature $T$ is proportional to the inverse of the radius
$r_\mathrm{H}$ of the apparent horizon and the radius $r_\mathrm{H}$ 
is proportional to the inverse of the Hubble rate $H$ \cite{Gibbons:1977mu}.
Therefore,  the temperature $T$ is proportional to the Hubble rate $H$. According to this statement, the behavior of $T$ or $H$ ca be used to test cosmological models nearby early and future singularities. Here we discussed in details these approach considering thermal corrections for quantum matter and classical fluids. The main result is that thermal effects can enhance or dissipate gravitational waves  depending on the sound speed of them into the cosmological medium and then depending on the equation of state. Clearly, if gravitational scalar modes are present, we have to consider an effective mass for the graviton \cite{FelixSergey} and this can be an indication to retain or discard  theories of gravity eventually modified with respect to GR. 
Since we are dealing with cosmology, it is difficult to identify astrophysical sources of gravitational waves so 
the stochastic background of gravitational waves could bring signatures of these phenomena 
which can be, in principle,  observationally probed.

Specifically, the stochastic background can be generated by 
the primordial background of gravitational waves or by the superposition of gravitational 
waves emitted by astrophysical objects that cannot be resolved \cite{Farmer:2003pa}.  In any case, a fundamental issue is to cross-correlate  astrophysical and cosmological gravitational wave backgrounds with respect to  the Cosmic Microwave Background as discussed in details in \cite {Ricciardone1}.

In the scenario discussed in this paper,  the stochastic background  is derived from the quantum fluctuations of 
zero-point energy acting in the primordial epochs. Such fluctuations are amplified in the early 
universe by the sudden variations of gravitational field. The mechanism produces a huge 
amount of gravitational waves and the theory results in agreement with an inflationary period 
and a spectral index of order 1. The PLANCK experiment data seem to confirm 
observationally this scenario \cite{Ade:2015rim}.

In general, 
the total stochastic background is described by a dimensionless spectrum 
\cite{Maggiore:1999vm,Grishchuk:2000gh}, that is 
\begin{equation}
\Omega^\mathrm{Tot}_\mathrm{GW}(\nu)=\frac{1}{\rho_{c}}
\left(\frac{d\rho^\mathrm{Tot}_\mathrm{GW}}{d\ln \nu}\right)\, ,\qquad
\label{eq: spettro}
\mbox{where}\quad
\rho_{c}\equiv\frac{3H_{0}^{2}}{8\pi G} \, .
\end{equation}
Here $\Omega^\mathrm{Tot}_\mathrm{GW}(\nu)$ is a dimensionless density parameter 
summing up all the gravitational contributions, $\rho_c$ is the cosmic critical energy density, 
$H_0$ the observed Hubble parameter at a given epoch, and
$d\rho^\mathrm{Tot}_\mathrm{GW}$ is the differential energy density of the
gravitational waves in the frequency range $\nu$ to
$\nu+d\nu$.

The total energy density, related to the stochastic background of gravitational waves, can be 
represented as $h_{0}^2 \, \Omega^\mathrm{Tot}_\mathrm{GW}(\nu)$ by
assuming a dimensionless Hubble parameter 
$H_0=100\,h_{0}\,\rm{km\, s^{-1}\,Mpc^{-1} }$. 
According to this definition, the stochastic background energy density can be given as a 
superposition of various components related to the propagation of gravitational waves. In our 
case, it is 
\begin{equation}
\Omega_\mathrm{GW}^\mathrm{Tot} \equiv \Omega_\mathrm{GW}^{+} 
+ \Omega_\mathrm{GW}^{\times} + \Omega_\mathrm{GW}^\mathrm{rad}
+ \Omega_\mathrm{GW}^S \, ,
\label{eq23}
\end{equation}
with $+,\times, \mathrm{rad}, S$ the indexes labeling the different modes.
The first two terms are the standard GR ones. The third is the enhancement or dissipation 
related to the thermal effects. The fourth represents any deviation from GR and, in particular, 
the scalar mode discussed in this paper.

It is worth stressing that inflation gives rise to perturbations for any 
tensor and scalar field. As a consequence, a spectrum of relic scalar and tensor 
gravitational waves is expected. In principle, such a spectrum can be a powerful testbed for 
any theory of gravity. 

It is possible to show that, at lower frequencies, the gravitational density parameter evolves 
as 
\begin{equation}
\Omega^\mathrm{Tot}_\mathrm{GW}(\nu)\propto \nu^{-2}\, .
\label{eq: spettro basse frequenze}
\end{equation} 
See, for example, \cite{Allen,Grishchuk:2000gh}. This feature can be important to probe 
radiation effects or scalar modes from observations. In particular, the 
characteristic amplitude for a given gravitational wave component is \cite{Maggiore:1999vm} 
\begin{equation}
h(\nu)\simeq8.93\times10^{-19}\left(\frac{1\,\mathrm{Hz}}{\nu}\right)
\sqrt{h_{100}^{2}\Omega_\mathrm{GW}(\nu)}\, ,
\label{eq: legame ampiezza-spettro}
\end{equation} 
where the amplitude is assumed at a strain frequency of $100$ Hz. This  quantity can be constrained by observations.

The constraints can be achieved by taking into account the
PLANCK experiment release and the LIGO-VIRGO operational frequencies \cite{ligo,virgo}. 
In the low frequency regime, we have 
\begin{equation} 
\Omega_\mathrm{GW}^\mathrm{GR}(\nu)h_{100}^{2}<2 \times
10^{-6}\,,\quad 
\Omega_\mathrm{GW}^\mathrm{rad}(\nu)h_{100}^{2}<2 \times
10^{-12}\,,\quad 
\Omega_\mathrm{GW}^{S}(\nu)h_{100}^{2}<2.3\times
10^{-12}\, ,
\label{eq: limite spettroGR}
\end{equation}
clearly indicating the differences in energy densities of tensor, radiation and scalar modes. 
This can be easily seen considering that $\rho_{GR}\sim H^2$, while 
$\rho_\mathrm{rad}\sim H^4$ as shown in \eqref{BRHR1}. 
Here, for the sake of simplicity, we are assuming 
$\alpha \sim 1$. In the same way, $\rho_S\sim H^4$ considering contributions of scalar 
fields in early epochs. See the discussion in Sec.~\ref{SecII}.

In particular, from Eq.~\eqref{TGW17}, thermal effects can be taken into account as soon as 
one succeeds in separating the first and second term in 
$h(t) \propto \e^{\left( 2 H_0 \pm H_0^2 \sqrt{\frac{\alpha}{2}} \right)t}$. 
Furthermore, considering the corresponding strain at 
$\approx 100{\,\mathrm{Hz}}$, for VIRGO and LIGO at
maximum sensitivity, we obtain 
\begin{equation}
h_{GR}(100\,\mathrm{Hz})<1.3\times 10^{-23}\,,\quad 
h_\mathrm{rad}(100\,\mathrm{Hz})<1.3\times 10^{-26}\,, \quad 
h_{S}(100\,\mathrm{Hz})<2\times 1.4 10^{-26}\,.
\label{eq: limite per lo strain}
\end{equation}
disentangling the amplitudes of GR, radiation term and scalar mode respectively. 

Clearly, the issue of separating thermal effects or scalar modes strictly depends on the 
maximum sensitivity at which the single interferometer (or the cluster of interferometers) is 
operating. In general, being $h_\mathrm{rad}\simeq h_{S}$ at early epochs, these last contributions could 
result difficult to separate.

For sensitivities of the order $10^{-22}$ of
the VIRGO and LIGO interferometers, at $\approx100\,\mathrm{Hz}$, one needs to gain two 
or three orders of magnitude. In the case of VIRGO, at a sensitivity of the order 
$10^{-21}$ at $\approx10\,\mathrm{Hz}$, it is 
\begin{equation}
h_{GR}(100\,\mathrm{Hz})<1.3 \times 10^{-22}\,,\quad h_\mathrm{rad}(100\,\mathrm{Hz})<1.3 
\times 10^{-25}\,, \quad h_{S}(100\,\mathrm{Hz})<2\times 10^{-25}\,. 
\label{eq: limite per lo strain01}
\end{equation} 
These qualitative results point out that tensor, radiation and scalar modes can be clearly 
distinguished considering suitable sensitivities of the instruments. More precisely, tensor 
modes and the global contribution of other modes can be disentangled as soon as suitable 
sensitivities are reached by interferometric systems (that is in one or more than one 
interferometers are working together, in particular, at least three correlated interferometers).

 However, gaining two or three orders of magnitude  in sensitivity from
earth-based interferometers is a very difficult task and  this could be  not sufficient to
detect gravitational waves stochastic background. Specifically, the probability of detection depends on the
algorithm which must be based on the output signal correlations \cite{Ricciardone2}. 
In this perspective, space-based interferometers can give satisfactory results. As discussed in \cite{Schmitz}, gravitational waves  produced by sound waves in the primordial  phase transitions are   a main target for the  Laser Interferometer Space Antenna (LISA) \cite{lisa}.  Our thermal effects can be included in this phenomenology. Furthermore, the LISA expected sensitivity for this type of gravitational  signals agrees with the order of magnitude necessary to disentangle standard GR modes with respect to gravitational scalar modes. In particular, being LISA  based on the concept of peak-integrated sensitivity curves \cite{Schmitz2}, it will be possible to  perform a systematic comparison of several thousands of benchmark points in different  models to obtain a complete information on the optimal signal-to-noise ratio. This procedure could allow, from one hand, to detect the stochastic background  and, from the other hand, to disentangle scalar modes.

Some concluding remarks are in order at this point. In this paper, we considered the effects 
of thermal radiation and scalar modes on the propagation of gravitational waves in dynamical 
cosmological backgrounds. Both kinds of effects can enhance or dissipate the total 
propagation of gravitational waves and represent significant signature to probe theories of 
gravity with respect to GR. Furthermore, also future singularities, as discussed above, can be 
affected by the thermal radiation or the presence of gravitational scalar modes. 

It is worth noticing that such effects give contributions also at quantum level so they can be 
related to primordial fluctuations of quantum matter. 

On the other hand, as shown in the case of $F(R)$ gravity, the effects of scalar modes or 
compressional fluids strictly depend on the ``representation'' of the theory in the Einstein or 
the Jordan frames. This could constitute an important feature in order to distinguish the true 
physical frame by the observations.

Finally, dynamics related to the above discussion could be observationally tested by 
interferometers. In fact, at suitable sensitivities, it seems realistic to disentangle GR 
contributions with respect to other contributions in the stochastic background of gravitational 
waves. In a future study, this topic will be developed in detail.

\begin{acknowledgments}
This work is partially supported by the JSPS Grant-in-Aid for Scientific Research (C) 
No. 18K03615 (S.N.). SC acknowledges the support of Istituto Nazionale di Fisica 
Nucleare (INFN), iniziative specifiche QGSKY and MOONLIGHT2 and 
SDO acknowledges Project No. PID2019-104397 GB-I00 from MINECO (Spain). 
\end{acknowledgments}

\end{document}